\documentclass[a4paper,10pt,twoside]{cpc-hepnp}

\usepackage{multicol}
\usepackage{graphicx}
\usepackage{booktabs}
\usepackage{amssymb,bm,mathrsfs,bbm,amscd}
\usepackage[tbtags]{amsmath}
\usepackage{lastpage}
\usepackage{subfigure}
\usepackage[justification=centering]{caption}

\begin{document}
\fancyhead[co]{\footnotesize ZHANG Yulian~ et al: Investigation of GEM-Micromegas detector on x-ray beam of synchrotron radiation}
\footnotetext[0]{Received 14 March 2012}
\title{Investigation of GEM-Micromegas detector on x-ray beam of synchrotron radiation \thanks{Supported by National Natural Science Foundation of China (11275224)}}
\author{%
      ¡SZHANG Yu-Lian$^{1,2,3}$%
\quad QI Hui-Rong$^{2,3;1)}$\email{qihr@ihep.ac.cn}\quad HU Bi-Tao$^{1}$ \quad FAN Sheng-Nan$^{2,3,4}$
\\WANG Bo$^{3}$ \quad LIU Mei$^{2,3,4}$ \quad ZHANG Jian$^{2,3}$ \quad LIU Rong-Guang $^{2,3}$
\\CHANG Guang-Cai$^{3}$ \quad LIU Peng$^{3}$ \quad OUYANG Qun$^{2,3}$ \quad CHEN Yuan-Bo$^{2,3}$
\\YI Fu-Ting$^{3}$
}
\maketitle
\address{%
$^1$ School of Nuclear Science and Technology, Lanzhou University, Lanzhou 730000, China\\
$^2$ State Key Laboratory of Particle Detection and Electronics,
Beijing 100049, China\\
$^3$ Institute of High Energy Physics, Chinese Academy of Sciences, Beijing 100049, China\\
$^4$ University of Chinese Academy of Sciences, Beijing 100049, China\\
}
\begin{abstract}
\noindent
  To reduce the discharge of the standard bulk Micromegas and GEM detector, the GEM-Micromegas detector was developed at the Institute of High Energy Physics. Taking into account the advantages of the two detectors, one GEM foil was set as a preamplifier on the mesh of Micromegas in the structure and the GEM preamplification decreased the working voltage of Micromegas to reduce the effect of the discharge significantly.
  At the same gain, the spark probability of the GEM-Micromegas detector can be reduced to a factor 0.01 compared to the standard Micromegas detector, and even the higher gain could
be obtained.
  In the paper, the performance of the detector in X-ray beam was studied at 1W2B Laboratory of Beijing Synchrotron Radiation Facility. Finally, the result of the energy resolution under various X-ray energies was given in different working gases. It indicates that the GEM-Micromegas detector has the energy response capability in the energy range from 6 keV to 20 keV and it could work better than the standard bulk-Micromegas.

\end{abstract}
\begin{keyword}
gaseous detector, energy resolution, synchrotron radiation
\end{keyword}

\begin{pacs}
29.40.Cs
\end{pacs}
\begin{multicols}{2}

\section{Introduction}

Micromegas (mi-cro-mesh-gaseous structure)\cite{lab1} is a widely used gaseous detector in high energy physics, which profits from its excellent position resolution and ability to work at high counting rates. However, one of the main problems is that it is particularly vulnerable to spark and discharge\cite{lab2} when the working voltage is higher. GEM (gas electron multiplier)\cite{lab3} is another kind of micro-pattern gaseous detector(MPGD)\cite{lab4} which has also been extensively researched and implemented in several large-scale high-energy physics experiments at CERN and elsewhere. GEM has many advantages such as brilliant position resolution, good energy resolution, toleration of high counting rate and easy assembly. Triple and quadruple GEM foils are usually cascaded to build higher gain detectors\cite{lab5}. The problem which arises from the cascaded GEM detector is that the last GEM foil (the foil near the readout plan) is quite easily damaged to spark because of the large number of electrons passing through it.

To decrease the spark probability, the GEM-Micromegas detector was proposed and tested in several applications of high energy physics\cite{lab6}\cite{lab7}. There was a GEM foil as a preamplifier element in the structure to reduce the amplification stress on Micromegas. The function of preamplification by the single GEM was ideal to obtain higher gain needed, but with a lower working voltage on Micromegas detector.
The assembly of GEM-Micromegas detector and its performance were described in the previous paper\cite{lab8}. At the same gain, the spark probability of the GEM-Micromegas detector can be reduced to a factor 0.01 compared to the standard Micromegas detector, and even the higher gain could be obtained. We reported the experiment carried out at the X-ray absorption station (Beamline 1W2B) of the Beijing Synchrotron Radiation Facility (BSRF) in January.
Compared with the position resolution of the Micromegas and GEM-Micromegas detectors, the energy resolution was less tested in different working gases and different energies of X-ray. In synchrotron radiation applications, there are demands for gaseous detectors with good position resolution as well as good energy resolution.
In this experiment, the energy resolution property of GEM-Micromegas detector was investigated to expand our preceding experiments and compared with that of the Micromegas detector. The aim of the present activity was to explore the possibility of using GEM-Micromegas detector in synchrotron radiation applications by analysing its energy resolution under different X-ray energies. 

\section{Experimental principle and setup}

\subsection{Principle of X-ray detection on synchronous radiation }

In the laboratory, three types of X-ray sources could be provided. They are the $^{55}$Fe radioactive source, the X-ray machine and the synchrotron radiation facility.
The $^{55}$Fe source could provide specific energy of X-ray but the intensity is low. To change the voltage and current of X-tube, higher energy and higher intensity of X-ray could be got by using X-ray machine, but the spectrum of X-ray is continuous. The synchrotron radiation facility not only provides the single energy of X-ray, but also has higher intensity in small beam spot.
The schematic diagram of SR beam test is illustrated in Fig.~\ref{fig0}.



\begin{center}
\includegraphics[width=6cm,height=3.6cm]{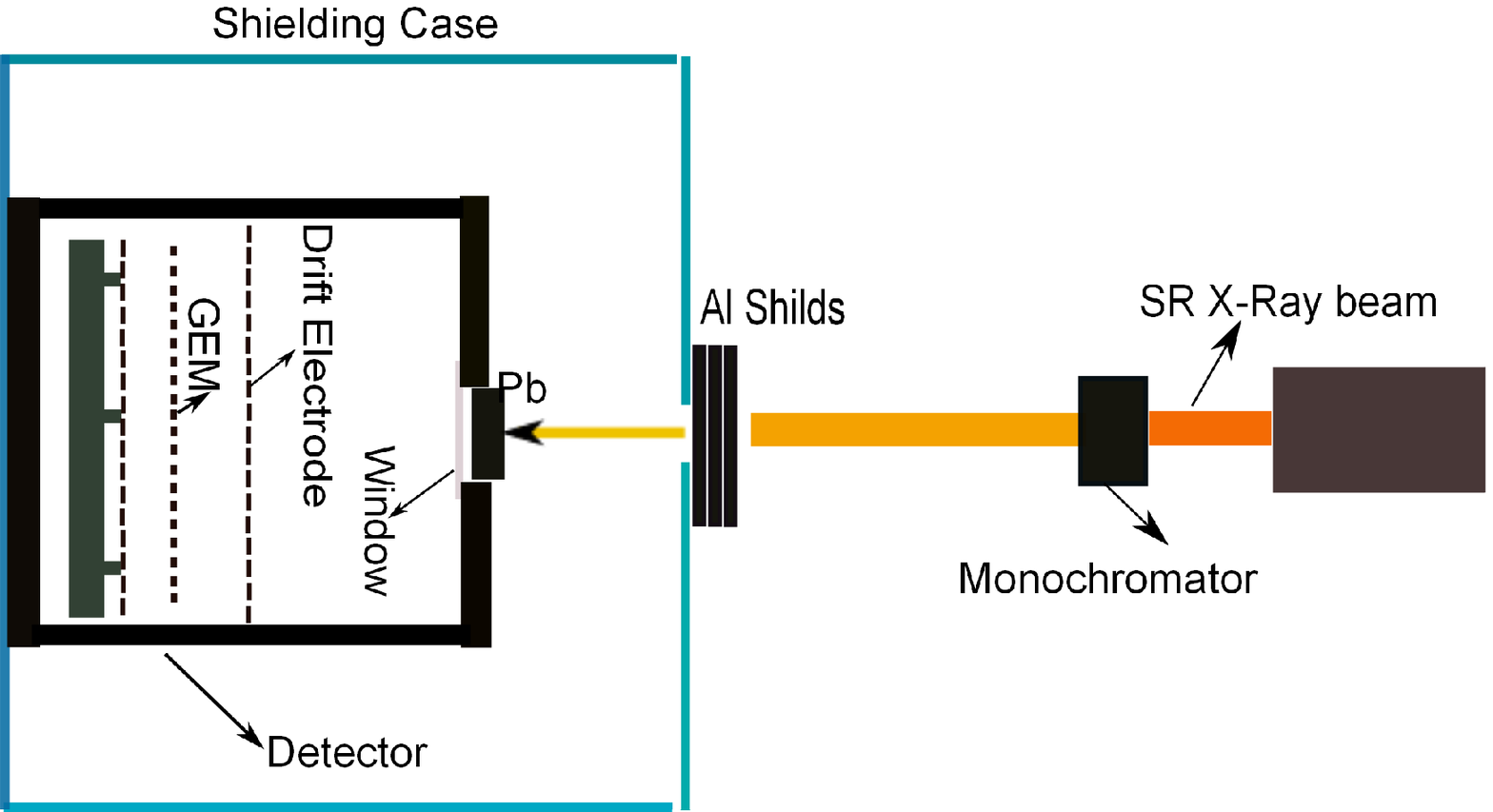}
\figcaption{\label{fig0} The schematic diagram of testing on BSRF.}
\end{center}
By adding a double crystal monochromator in the beam-line, the polychromatic light can be monochromatized and monochromatic light will be available. The intensity of Beamline 1W2B is ${10^{12}}$ photons/s and the spot size is $0.6 mm^2$ using the Pb collimator. To guarantee safe operation of the detector, the following steps are taken to protect the detector from too high X-ray intensity:

1) The beam stop window of Pb is assembled in the exit of beam spot and the thickness is 2.0mm.

2) Different numbers of layers of Al foils with thickness of 50um are stuck on the window of the detector to reduce the noise of background electromagnetic and  the intensity of X-ray.

 In the detector, the X-ray will interact with the working gas and produce primary ion-electron pairs in Region 1 or Region 2. It is shown in Fig.~\ref{fig1}. Under the drive of electric field, electrons emerging in Region 1 drift toward the read-out pads and get avalanched when they pass through the GEM foil and Region 3. The gain in the region1 is defined as $G_{GEM-MM}$ and the electron is avalanched by the GEM foil and the Micromegas detector.
\begin{center}
\includegraphics[width=5cm,height=3cm]{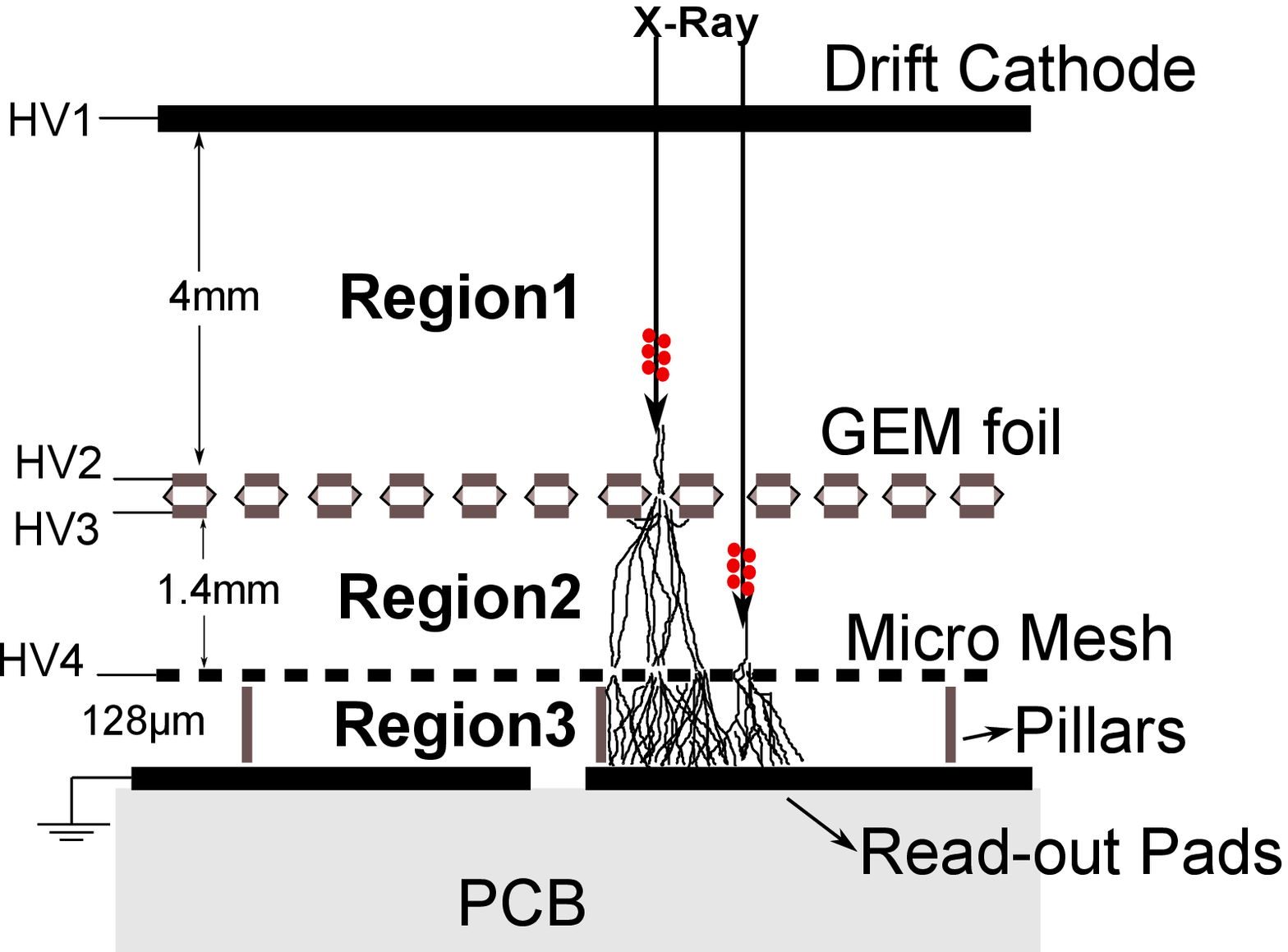}
\figcaption{\label{fig1} A schematic diagram of the prototype GEM-Micromegas detector.}
\end{center}
 In Region 2, the electron is just only avalanched by the Micromegas detector, the gain is defined as $G_{MM}$. Thus the gain resulting from the GEM can be calculated as:

\begin{equation}
\label{one}
\ G_{GEM}=\frac{G_{GEM-MM}}{G_{MM}}
\end{equation}
Compared with the standard Micromegas detector, the GEM-Micromegas detector shows the same or even higher gain under the identical working voltage on Micromegas. Thus the Micromegas could work under relatively lower voltage, thereby reducing the spark and discharge rate, which would ensure the security of the detector. Meanwhile, the disadvantage of the cascaded GEMs, which is the huge amount of electrons in the tiny hole of the last GEM foil make it the most vulnerable to sparking can be avoided.

The energy calibration of the beamline has been done, the principle is shown in Fig.~\ref{fig2}.
\begin{center}
\includegraphics[width=5.5cm,height=2.3cm]{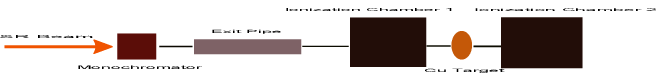}
\figcaption{\label{fig2} The test diagram of energy calibration of BSRF}
\end{center}
The attenuation of an X-ray beam traversing a medium could be written as:
\begin{equation}
\label{two}
\ \frac{I_{1}}{I_{2}}=e^{-\mu D}
\end{equation}
$I_{1}$ and $I_{2}$ denote the intensity of the beam traversing before and after the Cu target, recorded by Ionization 1 and Ionization 2, ${\mu}$ stands for the absorption coefficient, and D is the thickness. Then the absorption spectrum of Cu target could be calculated, it is $ \ln(I_{1}/I_{2}) $ as a function of X-ray energy. For example, the  X-ray  absorption  spectra  at  the  K-edge  of  copper is measured as 8977.766eV, and the value from the standard library is 8978.876eV. After calibration of the energy, the deviation of minus 1.1eV is obtained with the actual energy and the setting energy. It indicates that the value of X-ray's energy is very precise in 1W2B experimental station.

X-ray photons interact with matter in three different ways, and the photoelectric conversion is the dominant physical process in the experiment. Take Argon for example, the binding energy of the electrons in the shell i is $E_{i}$, the energy of the ejected photoelectron is:

\begin{equation}
\label{three}
\ E_{pe}=E_{X-ray}-E_{K}
\end{equation}

The excited atom could return to a lower state by emitting characteristic X-ray with energy $E_{X'-ray}$, or by transferring the energy to electrons within the atom and auger electron escaped carrying the excess energy $E_{Auger}$:

\begin{equation}
\label{four}
\begin{split}
& E_{X'-ray}=E_{K}-E_{L}\\
& E_{Auger}=E_{X'-ray}-E_{L}
\end{split}
\end{equation}

In energy spectrum, the values of full energy peak and escape peak could be calculated as\cite{lab9}:

\begin{equation}
\label{five}
\begin{split}
& E_{Full}=E_{pe}+E_{Auger}\\
& E_{Escape}=E_{pe}
\end{split}
\end{equation}

In the experiment, the energy spectrum is obtained under different X-ray energies. When X-ray with certain energy enters the detector, the values of the two energy peaks could be calculated by the theoretical functions. In the next section, the result of the comparison will be given with the experimental data.


\subsection{Experimental schemes}
The Micromegas detector is based on the bulk method and has the active area of $25mm\times25mm$.
The cascaded structure of the GEM-Micromegas detector is composed of the drift electrode, the standard GEM foil of $25mm\times25mm$ from CERN, the standard Micromegas of $25mm\times25mm$ by IHEP and the readout printed circuit board.
The signal is collected by the avalanche electrode and amplified by the charge sensitive preamplifier of ORTEC 142IH and the amplifier of ORTEC 572 A. All information of signal is acquired by the MCA of ORTEC ASPEC 927.
The high voltage supply of CAEN SY127 will provide the voltage independently for each electrode of the detector.
In the four mixture gases, both the GEM-Micromegas and the standard Micromegas have been tested. These mixture gases are: {90\%}Argon+{10\%}$CO_{2}$, {70\%}Argon+{30\%}$CO_{2}$, {95\%}Argon+{5\%}Isobutane and {90\%}Argon+{10\%}$CH_{4}$. The energy resolution of GEM-Micromegas detector is measured in the synchrotron radiation X-ray source and compared with that of the Micromegas detector.



\section{The experimental results and discussions}


\end{multicols}

\begin{figure}[!htbp]

\centering
\subfigure[]{
\label{fig:4:a} 
\includegraphics[width=7cm,height=4.4cm]{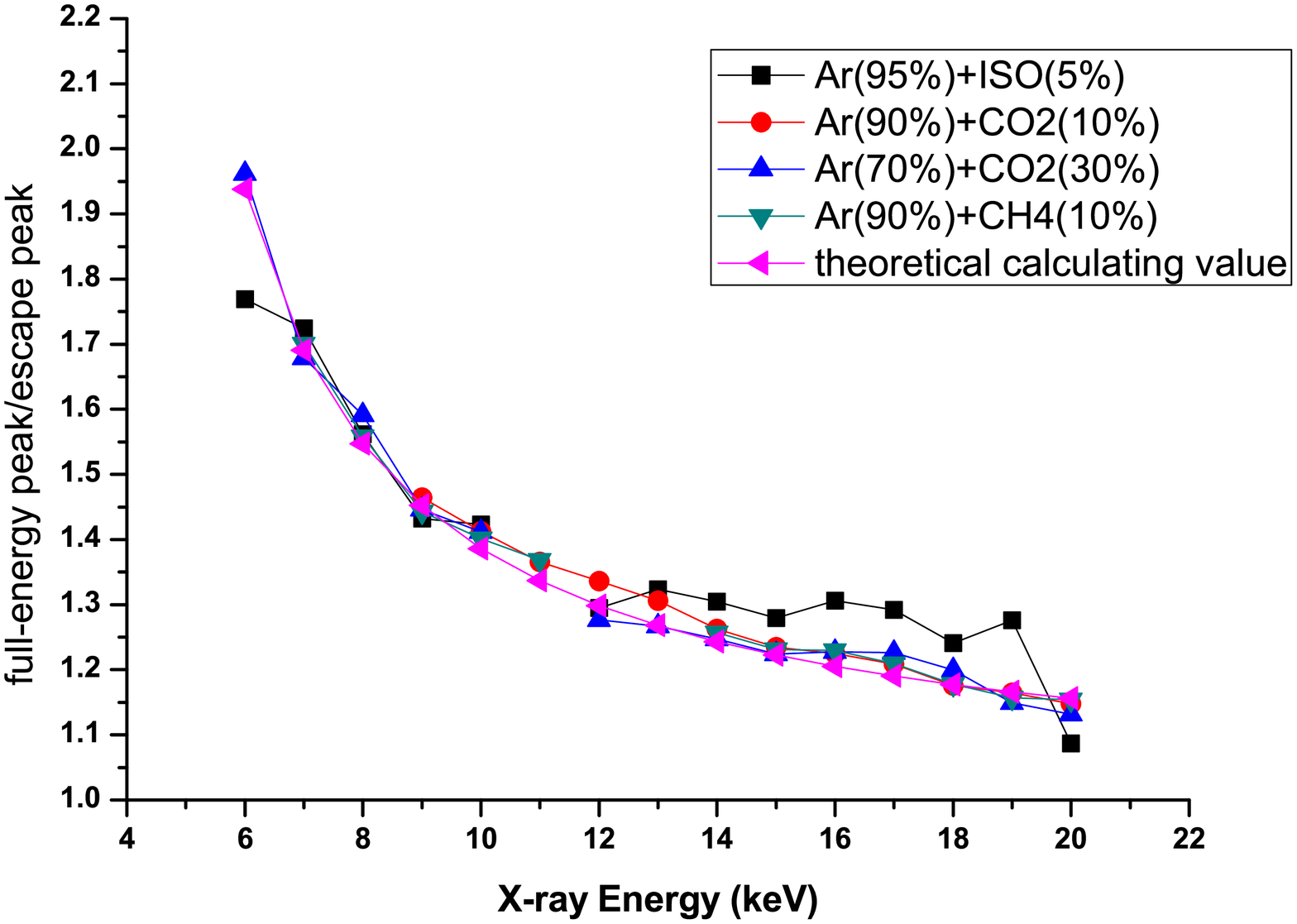}}
\hspace{0.5in}
\subfigure[]{
\label{fig:4:b} 
\includegraphics[width=7cm,height=4.4cm]{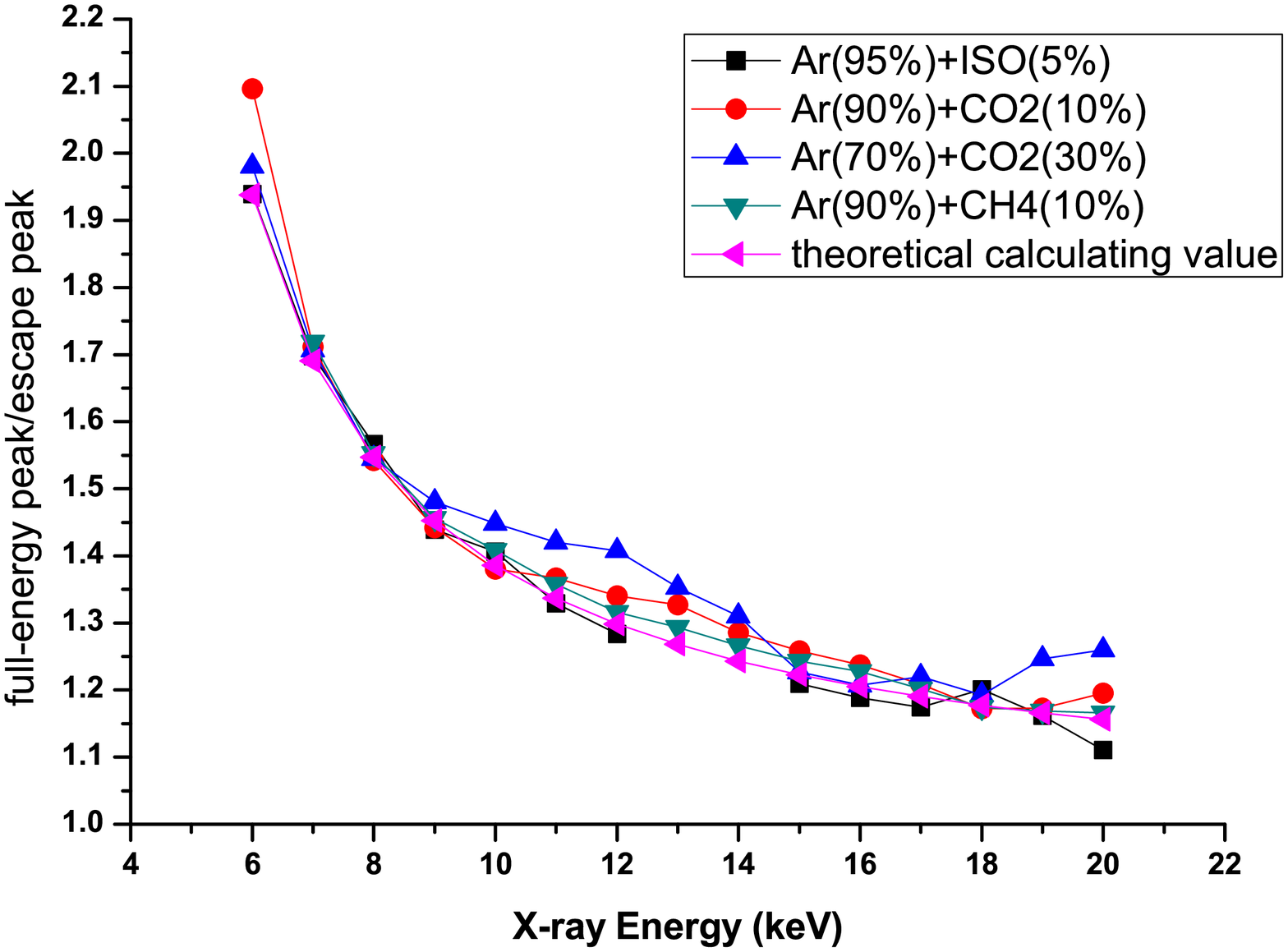}}
\hspace{0.5in}
\caption{The ratio of the full energy and escape peak in different energies and different working gases.\protect\\(Micromegas detector(a) and the GEM-Micromegas detector(b))}
\label{fig:3} 
\end{figure}

\begin{figure}[!htbp]
\begin{center}
\includegraphics[width=10cm,height=5.0cm]{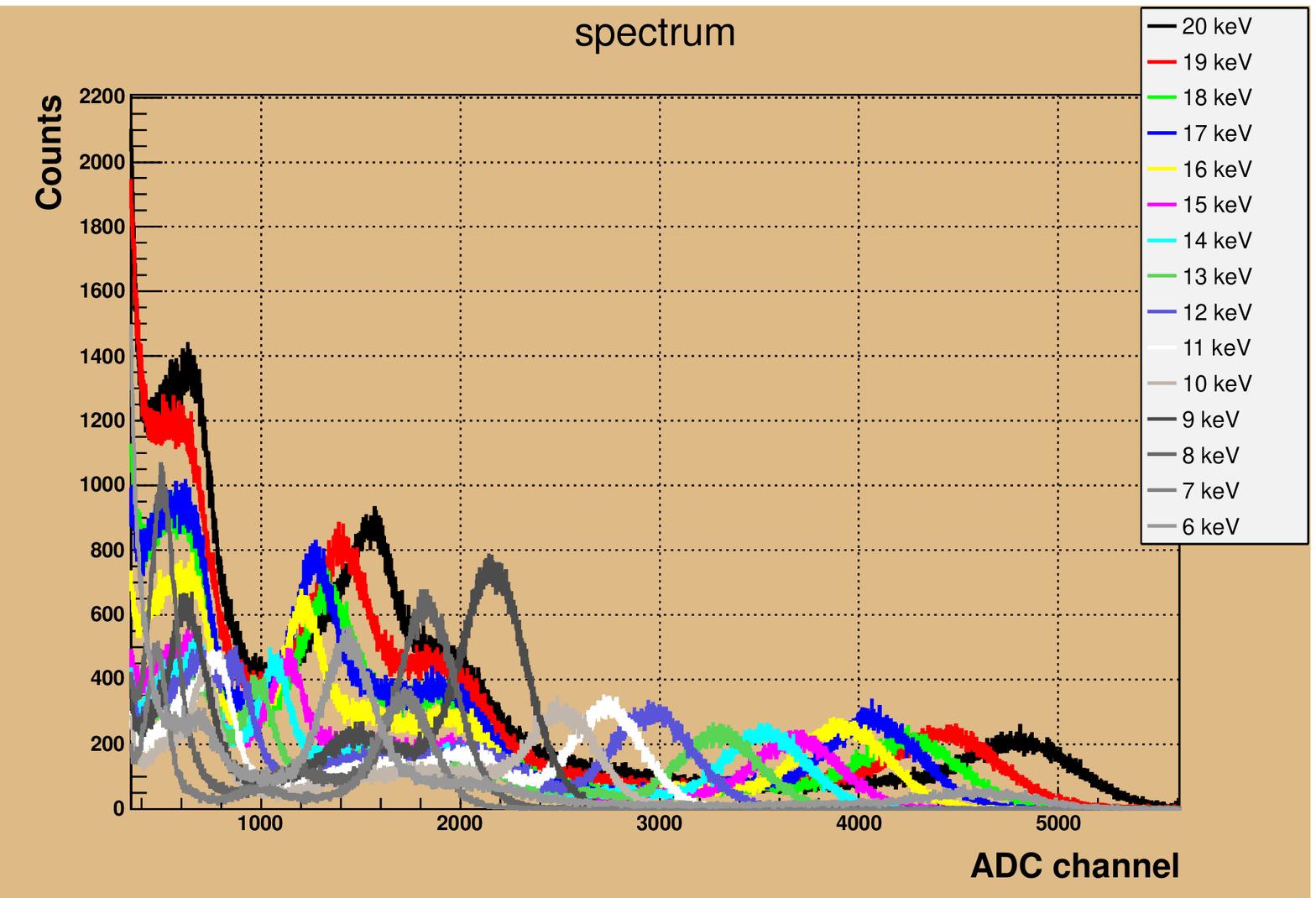}
\figcaption{\label{fig4} Energy spectrum under various X-ray energies.\protect\\(Working gas: {90\%} Argon, {10\%} $CO_{2}$)}
\end{center}
\end{figure}

\begin{multicols}{2}
In every specific working gas, the gain curve and energy resolution are first tested with a $^{55}$Fe X-ray source to determine the proper working voltage of the detectors.
Then the detectors are assembled in the X-ray Beam-line.
To ensure the detector works safely during the test, several steps are taken to protect the detectors from higher X-ray beam intensity.  In the experimental process, the beam stop of Pb layer is added in the center of beam spot, at the same time, several layers of aluminum foil are regarded as the attenuating material to reduce the X-ray's intensity. Changing a single energy of the synchrotron radiation X-ray, all of the corresponding energy spectrums will be obtained.  The value of full energy peak and escape peak are got by analyzing the energy spectrum and compared with the theoretical calculating value. The results are given in Fig.~\ref{fig:3}.
Theoretical calculating value is obtained according to Equations 3, 4 and 5. The range of values for $E_{X-ray}$ is from 6 keV to 20 keV, $E_{K}$ is 3.2 keV, and $E_{L}$ is 0.287 keV.
In different working gases and different X-ray energies, the results of the two energy peaks show good agreement with the experimental data and the theoretical data.
In the X-ray beam test, the energy spectrum under various X-ray energy is shown in Fig.~\ref{fig4}.
\end{multicols}
\begin{multicols}{2}

The results of energy resolution by Gaussian fitting of the full energy peak are shown in Fig 6. They indicated that both of the two detectors have energy resolutions which were better than {30\%}.
The GME-Micromegas detector could get more stable energy resolution at all energy range than the standard Micromegas detector. At higher energy, the GEM preamplification in the detector decreases the working voltage of Micromegas to reduce the affect of the discharge significantly. Over 8 keV, the energy resolution of GEM-micrpmeagas detector is less than {20\%} and it is better than Micromegas detector under the same test condition.

\end{multicols}
\begin{figure}[!htbp]

\centering
\subfigure[]{
\label{fig:6:a} 
\includegraphics[width=7cm,height=4.5cm]{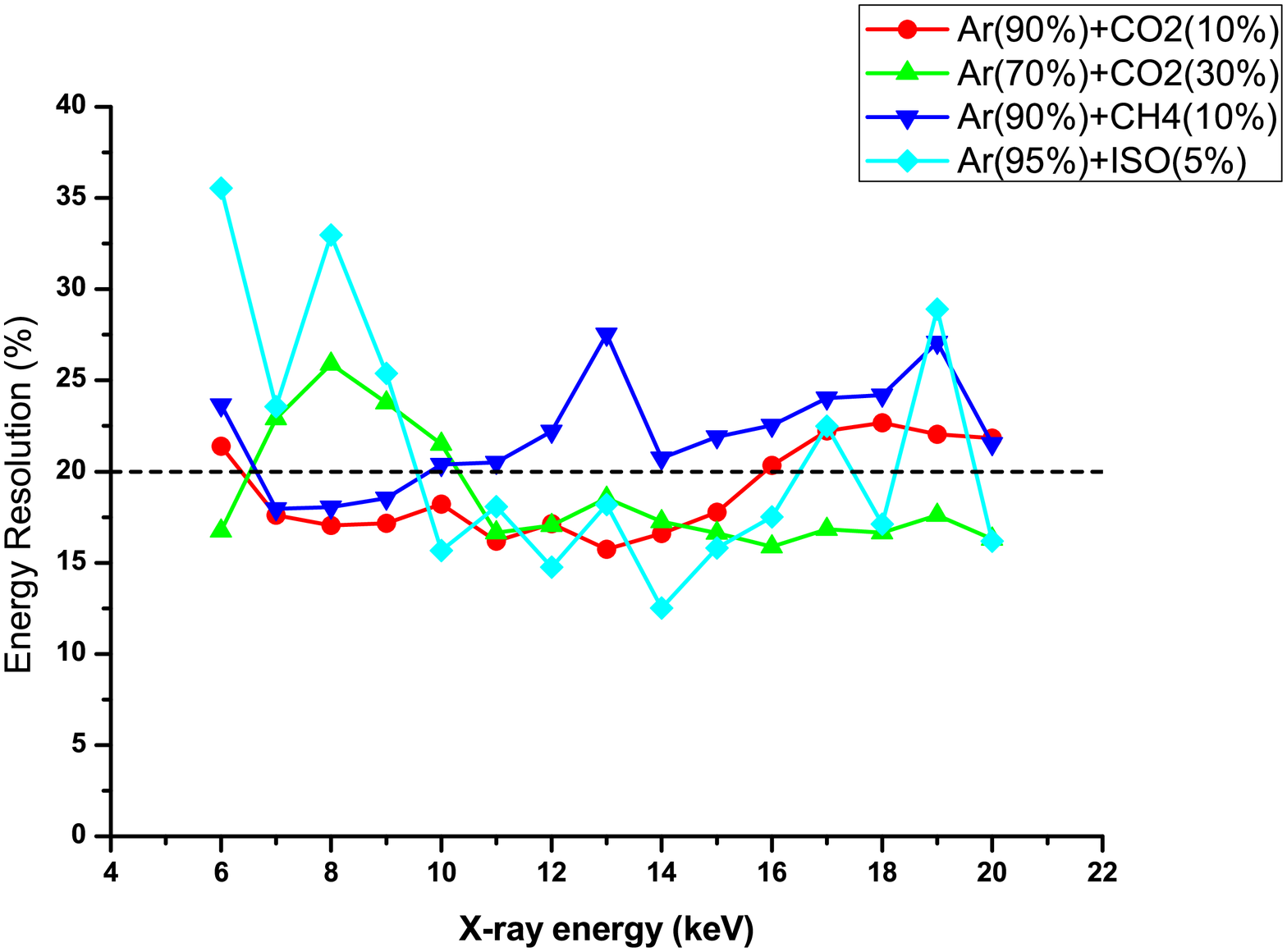}}
\hspace{0.5in}
\subfigure[]{
\label{fig:6:b} 
\includegraphics[width=7cm,height=4.5cm]{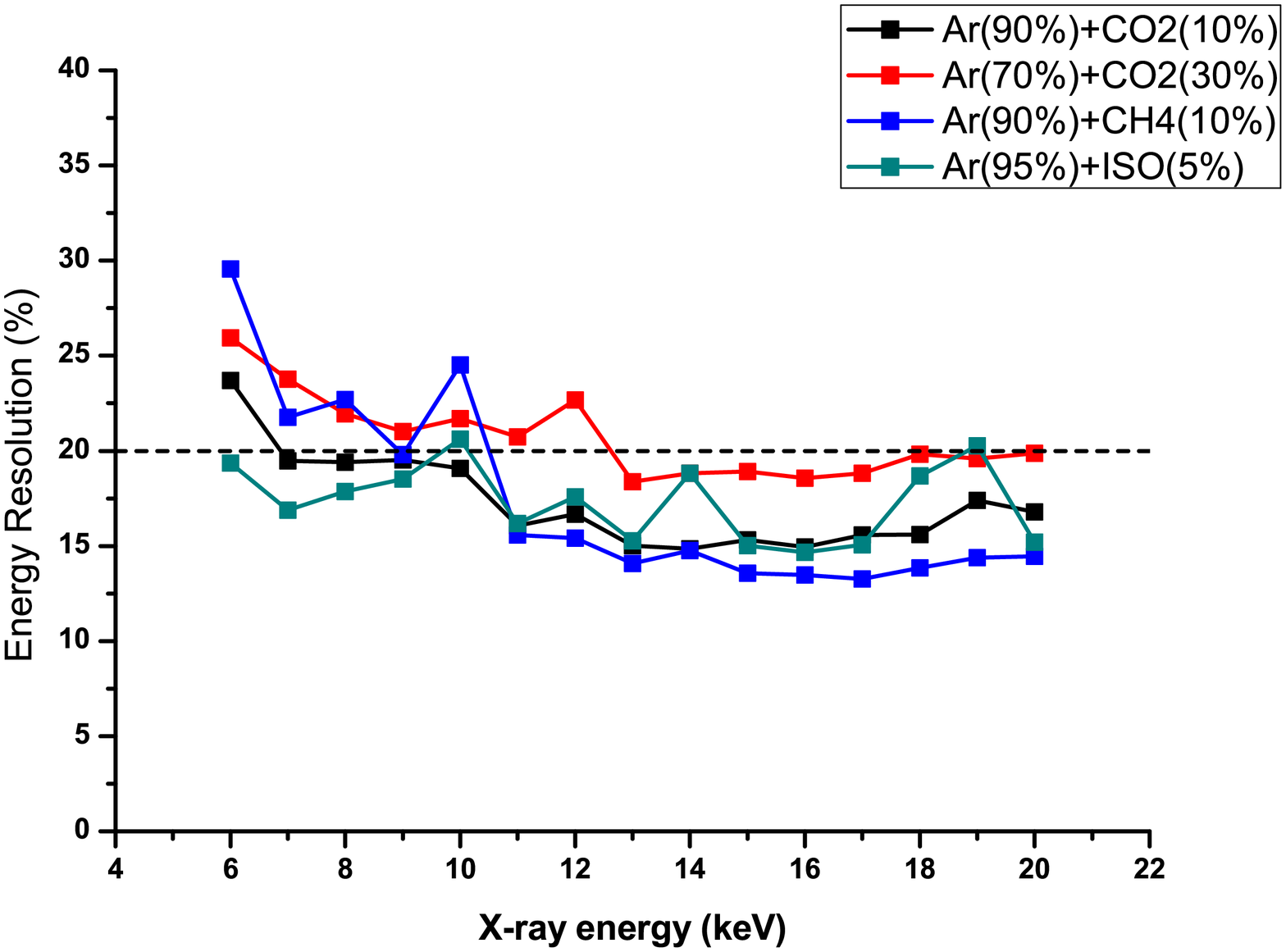}}
\hspace{0.5in}
\caption{Energy resolution of Micromegas(a) and GEM-Micromegas(b).}
\label{fig:6} 
\end{figure}


\begin{multicols}{2}

\section{Conclusion}

In four different mixed working gases and the X-ray energy range from 6 keV to 20 keV, the GEM-Micromegas detector has been tested in beam X-ray on BSRF. The energy spectrum and energy resolution are obtained. The results indicate that the GEM-Micromegas detector has the energy response capability in all the energy range and it could work better than the standard bulk-Micromegas.
Compared with standard Micromegas detector, the influence of discharge decreases and the GEM-Micromegas detector could maintain stable operation. Over 8 keV, the detector could obtain a better than {20\%} of energy resolution. It can meet part of the demand in synchrotron radiation research. In the next step, the size of the charge distribution requirs further measurement.
\section{Acknowledgments}
We are grateful to all colleagues at 1W2B Laboratory of Beijing Synchrotron Radiation Source. We wish to thank Tian Lichao, Xiu Qinglei and Liu Yi for their help and discussion.
\end{multicols}

\vspace{-1mm}
\centerline{\rule{80mm}{0.1pt}}
\vspace{2mm}

\begin{multicols}{2}

\end{multicols}

\clearpage

\end{document}